# Exact Solutions of Two-dimensional and Tri-dimensional Consolidation Equations

Romolo Di Francesco
GEO&GEO Instruments® - research & development
Teramo (TE), Italy
E-mail: romolo.difrancesco@vodafone.it

**Abstract.**
The exact solution of Terzaghi's consolidation equation has further highlighted the limits of this theory in the one-dimensional field as, like Taylor's approximate solution, it overestimates the decay times of the phenomenon; on the other hand, one only needs to think about the accumulation pattern of sedimentary-basin soils to understand how their internal structure fits in more with the model of transversely isotropic medium, so as to result in the development of two- and three-dimensional consolidation models. This is the reason why, using Terzaghi's theory and his exact solution as starting point, two-dimensional and three-dimensional consolidation equations have been proposed, in an attempt to find their corresponding exact solutions which constitute more reliable forecasting models. Lastly, results show how this phenomenon is predominantly influenced by the dimensions of the horizontal plane affected by soil consolidation and permeabilities that behave according to three coordinate axes.

*key words:* two-dimensional consolidation, three-dimensional consolidation, horizontal permeability, excess pore pressures, settlements, exact solution

## 1. Introduction

Terzaghi & Fröhlich's (1936) theory of consolidation, as it is limited to the one-dimensional field acting along the vertical direction alone, overestimates the decay times of the phenomenon; a limitation that had already become evident in Taylor's (1948) approximate solution and was further highlighted by the search for the exact solution (Di Francesco, 2011).
This limitation is overcome through the study of the dependence of hydraulic and mechanical soil behaviour on the effects of anisotropy - typically affecting sedimentary basins – which is due to internal structures, depending on the accumulation and deformation patterns in spatial and non-linear oedometer conditions; in other words, the geological evaluation of these basins generally leads to a structure that may more appropriately be described mathematically if it is treated as a transversely isotropic medium that is provided with a plane of horizontal isotropy and an axis of vertical symmetry. As these structures imply a coefficient of horizontal permeability that is greater than the vertical one:

$$K_h > K_v \tag{1}$$

the consolidation must necessarily be dependent on this condition, which may be extended also to the three-dimensional case and is more consistent with the actual hydraulic and mechanical behaviour of soils.
Starting from this assumption, this research initially illustrates two-/three-dimensional consolidation equations obtained from the continuity equation of saturated soils; then the focus is on their exact solution. Finally, this is completed with practical applications that are able to highlight the dependence of the mechanical response of soils on oedometer deformations on the condition imposed by the eq. (1).

## 2. Two- and Three-dimensional Consolidation Equations

The starting point is the continuity equation valid for saturated soils which, in the general case, takes the following form (Lambe & Whitman, 1969):

$$\left(K_x \cdot \frac{\partial^2 h}{\partial x^2}\right) + \left(K_y \cdot \frac{\partial^2 h}{\partial y^2}\right) + \left(K_z \cdot \frac{\partial^2 h}{\partial z^2}\right) = \frac{1}{1+e} \cdot \frac{\partial e}{\partial t} \qquad (2)$$

being expressed as a function of the piezometric head $h$ and having assumed the invariability of the coefficients of permeability.

The definition of the consolidation law requires the introduction in the eq. (2) of the indefinite equations of soil balance and of the relationship of stress-strain and deformation of the solid skeleton in oedometer conditions. According to the pattern in Lambe & Whitman (1969) for the one-dimensional case, this leads to the following formulation that is valid in the three-dimensional field:

$$c_{h(x)} \cdot \frac{\partial^2 u}{\partial x^2} + c_{h(y)} \cdot \frac{\partial^2 u}{\partial y^2} + c_v \cdot \frac{\partial^2 u}{\partial z^2} = \frac{\partial u}{\partial t} \qquad (3)$$

where the horizontal coefficient of consolidation:

$$c_{h(x)} = \frac{K_x}{\gamma_w \cdot m_{h(x)}} \qquad (4)$$

$$c_{h(y)} = \frac{K_y}{\gamma_w \cdot m_{h(y)}} \qquad (5)$$

as well as the vertical coefficient appear:

$$c_v = \frac{K_z}{\gamma_w \cdot m_v} \qquad (6)$$

Lastly, in the case of two-dimensional filtration the eq. (3) is reduced to:

$$c_h \cdot \frac{\partial^2 u}{\partial x^2} + c_v \cdot \frac{\partial^2 u}{\partial z^2} = \frac{\partial u}{\partial t} \qquad (7)$$

thus showing, similarly to three-dimensional consolidation, the dependence on the condition imposed by the eq. (1) that may translate in $c_h > c_v$.

## 3. Exact Solution of the Two-dimensional Consolidation

The eq. (7) may be written in another, fully equivalent, form as a function of the pore pressure variation:

$$c_h \cdot \frac{\partial^2 \Delta u}{\partial x^2} + c_v \cdot \frac{\partial^2 \Delta u}{\partial z^2} = \frac{\partial \Delta u}{\partial t} \qquad (8)$$

whose solution:

$$\Delta u(x,z,t) = \Delta u \cdot e^{-k_{u(x)} \cdot x} \cdot e^{-k_{u(z)} \cdot z} \cdot \cos\left(\omega \cdot t - k_{u(x)} \cdot x - k_{u(z)} \cdot z\right) \qquad (9)$$

may be derived by extending the exact solution of the one-dimensional equation (Di Francesco R., 2011) in relation to the consolidation variables $k_{u(x)}$ e $k_{u(z)}$.

Likewise, the validity of the eq. (9) requires the calculation of the first derivative as against time:

$$\frac{\partial \Delta u}{\partial t} = -\omega \cdot \Delta u_0 \cdot e^{-k_{u(x)} \cdot x} e^{-k_{u(z)} \cdot z} \cdot \sin\left(\omega \cdot t - k_{u(x)} \cdot x - k_{u(z)} \cdot z\right) \qquad (10)$$

and of the second derivative as against *z*:

$$\frac{\partial \Delta u}{\partial z} = -k_{u(z)} \cdot \Delta u \cdot e^{-k_{u(x)} \cdot x} \cdot e^{-k_{u(z)} \cdot z} \cdot \cos\left(\omega \cdot t - k_{u(x)} \cdot x - k_{u(z)} \cdot z\right) + \qquad (11)$$
$$+ k_{u(z)} \cdot \Delta u \cdot e^{-k_{u(x)} \cdot x} \cdot e^{-k_{u(z)} \cdot z} \cdot \sin\left(\omega \cdot t - k_{u(x)} \cdot x - k_{u(z)} \cdot z\right)$$

$$\frac{\partial^2 \Delta u}{\partial z^2} = k_{u(z)}^2 \cdot \Delta u \cdot e^{-k_{u(x)} \cdot x} \cdot e^{-k_{u(z)} \cdot z} \cdot \cos\left(\omega \cdot t - k_{u(x)} \cdot x - k_{u(z)} \cdot z\right) - \qquad (12)$$
$$+ k_{u(z)}^2 \cdot \Delta u \cdot e^{-k_{u(x)} \cdot x} \cdot e^{-k_{u(z)} \cdot z} \cdot \sin\left(\omega \cdot t - k_{u(x)} \cdot x - k_{u(z)} \cdot z\right) -$$
$$+ k_{u(z)}^2 \cdot \Delta u \cdot e^{-k_{u(x)} \cdot x} \cdot e^{-k_{u(z)} \cdot z} \cdot \sin\left(\omega \cdot t - k_{u(x)} \cdot x - k_{u(z)} \cdot z\right) -$$
$$+ k_{u(z)}^2 \cdot \Delta u \cdot e^{-k_{u(x)} \cdot x} \cdot e^{-k_{u(z)} \cdot z} \cdot \cos\left(\omega \cdot t - k_{u(x)} \cdot x - k_{u(z)} \cdot z\right)$$

Hence, simplifying the eq. (12) and extending the calculation also as against *x* you obtain:

$$\frac{\partial^2 \Delta u}{\partial z^2} = -2 k_{u(z)}^2 \cdot \Delta u \cdot e^{-k_{u(x)} \cdot x} \cdot e^{-k_{u(z)} \cdot z} \cdot \sin\left(\omega \cdot t - k_{u(x)} \cdot x - k_{u(z)} \cdot z\right) \qquad (13)$$

$$\frac{\partial^2 \Delta u}{\partial x^2} = -2 k_{u(x)}^2 \cdot \Delta u \cdot e^{-k_{u(x)} \cdot x} \cdot e^{-k_{u(z)} \cdot z} \cdot \sin\left(\omega \cdot t - k_{u(x)} \cdot x - k_{u(z)} \cdot z\right) \qquad (14)$$

Lastly, by introducing the equations (13) and (14) in (8) and through simplification you obtain:

$$c_h \cdot k_{u(x)}^2 + c_v \cdot k_{u(z)}^2 = \frac{\omega}{2} \qquad (15)$$

For the solution of the eq. (15) it must be considered that from the eq. (1) the relationship between the coefficients of permeability applies:

$$\frac{K_x}{K_z} = m \qquad \rightarrow \qquad K_x = m \cdot K_z \qquad (16)$$

and, in a wider sense, between the coefficients of consolidation:

$$\frac{c_h}{c_v} = m \qquad \rightarrow \qquad c_h = m \cdot c_v \qquad (17)$$

and between the consolidation variables:

$$\frac{k_{u(x)}}{k_{u(z)}} = m \qquad \rightarrow \qquad k_{u(x)} = m \cdot k_{u(z)} \qquad (18)$$

By performing the required substitutions, and with a few mathematical manipulations, the eq. (15) becomes:

$$c_v \cdot k_{u(z)}^2 \cdot \left(m^3 + 1\right) = \frac{\omega}{2} \qquad (19)$$

from which the expression of the consolidation variable in vertical direction may be obtained:

$$k_{u(z)} = \sqrt{\frac{\pi}{c_v \cdot t \cdot (m^3 + 1)}} \qquad (20)$$

as the relationship $\omega = 2\pi/t$ applies.

The introduction of the eq. (20) in the eq. (18) results, in succession, in the definition of the consolidation variable in horizontal direction:

$$k_{u(x)} = m \cdot k_{u(z)} = m \cdot \sqrt{\frac{\pi}{c_v \cdot t \cdot (m^3 + 1)}} \qquad (21)$$

Please note that for the condition $m = 1$ equations (20) and (21) are reduced to:

$$k_{u(x)} = k_{u(z)} = \sqrt{\frac{\pi}{2 \cdot c_v \cdot t}} \qquad (22)$$

and that the consolidation variables must not be determined, as they are dependent upon the coefficients of consolidation that experimentally may be derived from oedometer tests.

## 4. Exact Solution of the Three-dimensional Consolidation

By using the same procedure illustrated for the two-dimensional consolidation, the eq. (3) may be written as:

$$c_{h(x)} \cdot \frac{\partial^2 \Delta u}{\partial x^2} + c_{h(y)} \cdot \frac{\partial^2 \Delta u}{\partial y^2} + c_v \cdot \frac{\partial^2 \Delta u}{\partial z^2} = \frac{\partial \Delta u}{\partial t} \qquad (23)$$

whose solution is:

$$\Delta u(x, y, z, t) = \Delta u \cdot e^{-k_{u(x)} \cdot x} \cdot e^{-k_{u(y)} \cdot y} \cdot e^{-k_{u(z)} \cdot z} \cdot \cos\left(\omega \cdot t - k_{u(x)} \cdot x - k_{u(y)} \cdot y - k_{u(z)} \cdot z\right) \qquad (24)$$

This is followed by the calculation of the first derivative as against time and of the second derivatives as against space, which, when replaced in the eq. (24), give:

$$c_{h(x)} \cdot k_{u(x)}^2 + c_{h(y)} \cdot k_{u(y)}^2 + c_v \cdot k_{u(z)}^2 = \frac{\omega}{2} \qquad (25)$$

Again, the solution of the eq. (25) is derived from the relationship between the coefficients of permeability:

$$\frac{K_y}{K_z} = n \quad ; \quad \frac{K_x}{K_z} = m \qquad (26)$$

between the coefficients of consolidation:

$$\frac{c_{h(y)}}{c_v} = n \quad ; \quad \frac{c_{h(x)}}{c_v} = m \qquad (27)$$

and between the consolidation variables:

$$\frac{k_{u(y)}}{k_{u(z)}} = n \quad ; \quad \frac{k_{u(x)}}{k_{u(z)}} = m \qquad (28)$$

which transform the eq. (25) in:

$$c_v \cdot k_{u(z)}^2 \cdot (m^3 + n^3 + 1) = \frac{\omega}{2} \tag{29}$$

from which the consolidation variable in vertical direction may be obtained:

$$k_{u(z)} = \sqrt{\frac{\pi}{c_v \cdot t \cdot (m^3 + n^3 + 1)}} \tag{30}$$

Here too, the introduction of the eq. (30) in the equations (28) results in the definition of the consolidation variables in the horizontal directions:

$$k_{u(x)} = m \cdot k_{u(z)} = m \cdot \sqrt{\frac{\pi}{c_v \cdot t \cdot (m^3 + n^3 + 1)}} \tag{31}$$

$$k_{u(y)} = n \cdot k_{u(z)} = n \cdot \sqrt{\frac{\pi}{c_v \cdot t \cdot (m^3 + n^3 + 1)}} \tag{32}$$

Lastly, for the symmetry properties of a natural deposit that may be identified as a transversely isotropic medium the following conditions apply:

$$K_x = K_y \tag{33}$$

$$k_{u(x)} = k_{u(y)} \tag{34}$$

which simplify the equations (30), (31), and (32) in:

$$k_{u(z)} = \sqrt{\frac{\pi}{c_v \cdot t \cdot (2m^3 + 1)}} \tag{35}$$

$$k_{u(x)} = k_{u(y)} = m \cdot \sqrt{\frac{\pi}{c_v \cdot t \cdot (2m^3 + 1)}} \tag{36}$$

To conclude, for the condition $m = n = 1$ the equations (31), (32), (35), and (36) are reduced to:

$$k_{u(x)} = k_{u(y)} = k_{u(z)} = \sqrt{\frac{\pi}{3 \cdot c_v \cdot t}} \tag{37}$$

where, at the same time, the same considerations apply as seen in the 2D case concerning the dependence of consolidation variables on the coefficients of consolidation.

## 5. Practical Applications

By limiting study scope on the two-dimensional consolidation alone, the starting point is given by the action of a static load N = 100 kPa, derived from the same example used in Di Francesco (2011), where application times may be neglected compared with clay filtration times; this evidences an initially undrained system that is capable of developing a theoretical excess pore pressure Δu = N = 100 kPa, in relation to the incompressibility of both solid skeleton and pore fluid (fig. 1a). A system that, when limited to the one-dimensional consolidation, shows dissipation times $t_{95\%}$ = 15.5 years according to Taylor's approximate solution (1948) and $t_{100\%}$ = 17.5 years according to Di Francesco's (2011) exact solution, the latter resolved both through the construction of "n" dissipation curves and the search for the analytical solution for the condition $\Delta u(z, t_{100})$ = 0 kPa (fig. 1b).

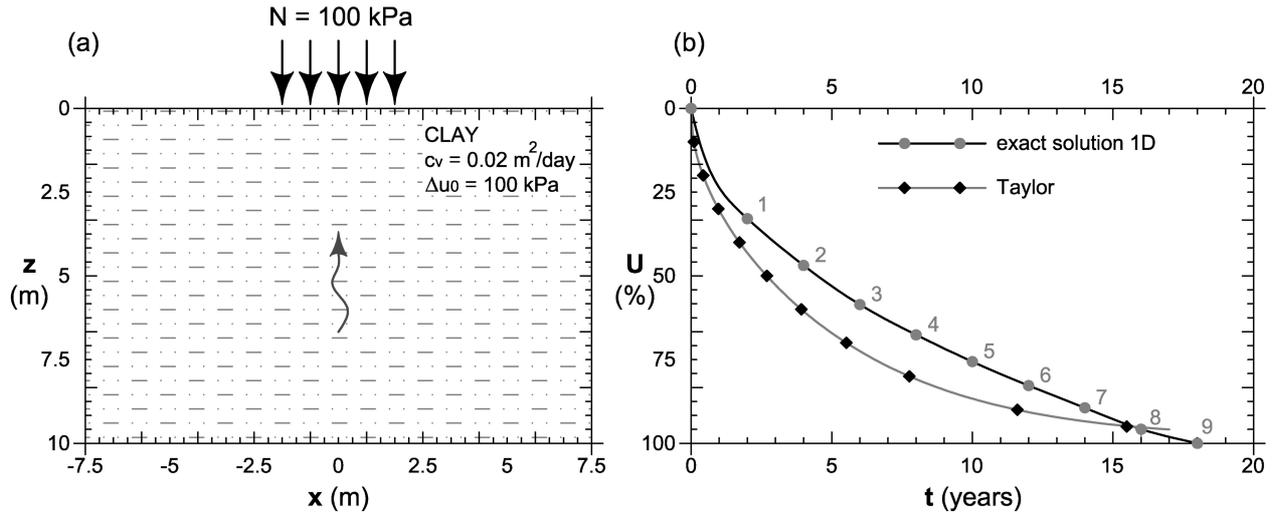

Fig. 1. Application example of the 2D consolidation's exact solution: (a) reference stratigraphy; (b) comparison between Taylor's approximate solution and the exact solution of Terzaghi's equation (by Di Francesco, 2011).

While the exact solution of the 1D consolidation finds immediate application, in the 2D consolidation the need arises to identify in advance any relationships existing between the coefficients of permeability (eq. (16)) and between the vertical and horizontal dimensions introduced in the eq. (9); as a result, in order to evaluate the influence of these parameters on the dissipation times of excess pore pressures, and the relevant development of oedometer settlements, the analyses were carried out attributing to "m" the values 1, 2, 5, and 10 and to "x" the values 0.1z, 0.5z, z, and 2z, as the dimensions of the portion of the layer subject to consolidation are not known *a priori*.

| Table 1. Decay times of 2D consolidation when varying "m" and "x" | | | | | | | | |
|---|---|---|---|---|---|---|---|---|
| z (m) | x (m) | m ($K_x/K_z$) | $c_v$ (m²/y) | $c_h$ (m²/y) | $k_{u(z)}$ (m⁻¹) | $k_{u(x)}$ (m⁻¹) | t (days) | t (years) |
| 10 | 1 (0.1z) | 1 | 0.02 | 0.02 | 0.1425 | 0.1425 | 3865.2 | 10.59 |
| 10 | 1 (0.1z) | 2 | 0.02 | 0.04 | 0.1306 | 0.2613 | 1022.2 | 2.80 |
| 10 | 1 (0.1z) | 5 | 0.02 | 0.1 | 0.1045 | 0.5225 | 114.1 | 0.31 |
| 10 | 1 (0.1z) | 10 | 0.02 | 0.2 | 0.0784 | 0.7838 | 25.53 | 0.07 |
| 10 | 5 (0.5z) | 1 | 0.02 | 0.02 | 0.1045 | 0.1045 | 7187.3 | 19.69 |
| 10 | 5 (0.5z) | 2 | 0.02 | 0.04 | 0.0784 | 0.1568 | 2839.5 | 7.78 |
| 10 | 5 (0.5z) | 5 | 0.02 | 0.1 | 0.0448 | 0.2239 | 621.1 | 1.70 |
| 10 | 5 (0.5z) | 10 | 0.02 | 0.2 | 0.0261 | 0.2613 | 229.8 | 0.63 |
| 10 | 10 (z) | 1 | 0.02 | 0.02 | 0.0784 | 0.0784 | 12777.7 | 35.01 |
| 10 | 10 (z) | 2 | 0.02 | 0.04 | 0.0523 | 0.1045 | 6388.9 | 17.50 |
| 10 | 10 (z) | 5 | 0.02 | 0.1 | 0.0261 | 0.1306 | 1825.4 | 5.00 |
| 10 | 10 (z) | 10 | 0.02 | 0.2 | 0.0143 | 0.1425 | 2.12 | 2.12 |
| 10 | 20 (2z) | 1 | 0.02 | 0.02 | 0.0523 | 0.0523 | 28749.9 | 78.77 |
| 10 | 20 (2z) | 2 | 0.02 | 0.04 | 0.0314 | 0.0627 | 17746.8 | 48.62 |
| 10 | 20 (2z) | 5 | 0.02 | 0.1 | 0.0143 | 0.0713 | 6135.3 | 16.81 |
| 10 | 20 (2z) | 10 | 0.02 | 0.2 | 0.0075 | 0.0746 | 2814.7 | 7.71 |

The results, summarised in Table 1, show that the more "m" increases, the lower consolidation times become for every value adopted for "x", in line with the behaviour expected for a two-dimensional medium in relation to the conditions expressed by the equations (16), (17), and (18); simultaneously, the more "x" increases, the higher the decay times of the phenomenon become, which may be attributed to the increase in the drainage path.

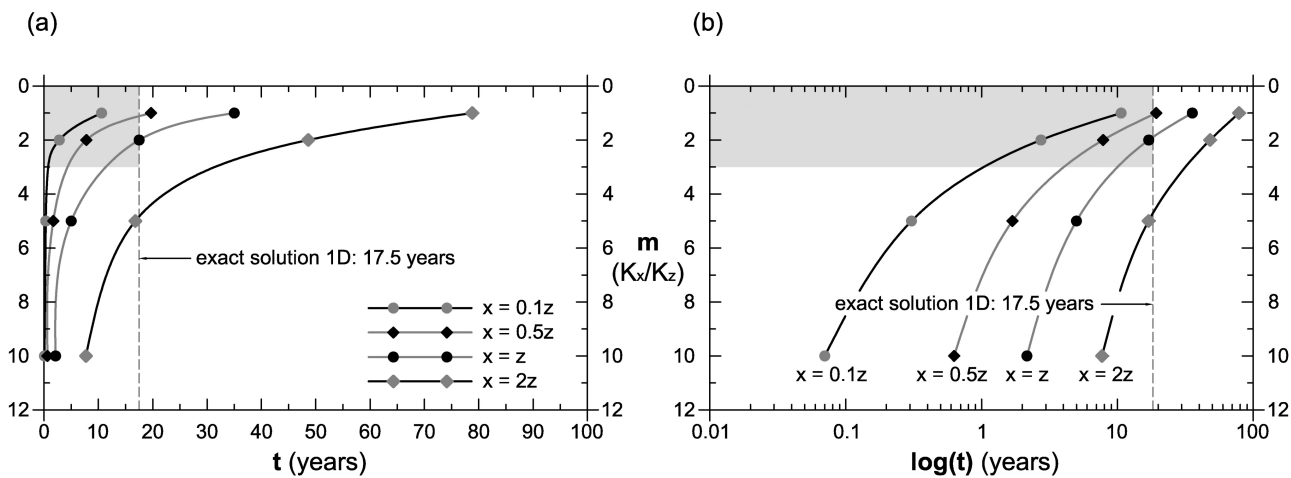

Fig. 2. Decay times of 2D consolidation related to fig. 1a, when varying "m" and "x", represented in linear (a) and semi-logarithmic (b) scale; the gray area identifies the possible values of "x" on the assumption that the experimental data m = 3 applies and in the presence of the condition $t_{2D} < t_{1D}$.

A graph of the results in the plane "m ÷ t" (fig. 2a) and in the plane "m ÷ log(t)" (fig. 2b) provides an insight of the influences exercised by both m" and "x", especially if they are compared with the decay time calculated with the exact solution of the 1D consolidation (fig. 1b) that, therefore, must represent a higher limit; a condition that, when combined with the experimental value of "m", allows to define the scope of the most likely value of "x", as is the case with the gray area in fig. 2 concerning a hypothetical m = 3. In other words, for a horizontal layer that is endlessly extended in this direction, the value of "x" to be included in the exact solution must depend on the foundation dimension and the consequent tensional level induced by the underground.

# 6. Conclusions

The differential equation pertaining to one-dimensional consolidation, in spite of its unquestionable conceptual validity, overestimates the decay times of the phenomenon so as to be extended to two- and three-dimensional fields; the application of the method used for the solution of Terzaghi's equation (Di Francesco, 2011) then helped identify its exact solutions which, as well as the consolidation variables $k_{u(x)}$, $k_{u(y)}$, and $k_{u(z)}$, include their relationships "m" and "n" and the drainage dimensions contained in the horizontal plane.

Lastly, based on the analysis of the application of the exact solution from the 2D consolidation, it was found out that, while consolidation variables do not need to be determined, as they are dependent upon the coefficients of consolidation, the decay times of the phenomenon are strongly conditioned by the horizontal dimension. This, in turn, must necessarily depend on the dimensions of the geotechnical structure under examination; an issue that may be clarified by retrospective review of any consolidation settlements monitored in real foundations.

In conclusion, it should also pointed out that, notwithstanding the dependence of the exact solutions on the coefficients of consolidation related to the horizontal plane ($c_{h(x)}$ and $c_{h(y)}$), which in turn are a function of the relevant compressibility ratios ($m_{h(x)}$ and $m_{h(y)}$), their application is actually simplified by the search for the relationships "m" and "n" that only depend on the permeabilities acting according to three coordinated directions; these elements may be easily determined through laboratory testing, both in oedometer and triaxial cells by rotating the axes when pulling out the test pieces from the samples to be analysed.